# Towards an Innate Cell-Environment Nanothermometer.

Cristinas Carrizo[1,2, ||], Gianluca D´Agostino[3, ||], Graham Spicer[4,5], Jaime Fernández de Córdoba[3], Rubén Ahijado Guzmán[6], Clara Maria Garcia-Abad[1,2], Aitor Rivas[1,2], Ruth Matesanz[7], Ana Oña[3] and Sebastian A. Thompson[1,2]*

_________

1 Madrid Institute for Advanced Studies in Nanoscience (IMDEA Nanociencia), Madrid, Spain.

2 Nanobiotechnology Unit Associated to the National Center for Biotechnology (CNB-CSIC-IMDEA), Madrid, Spain.

3 Advanced Light Microscopy, Centro Nacional de Biotecnología (CNB-CSIC), Madrid, Spain.

4 Wellman Center for Photomedicine, Massachusetts General Hospital, Boston, MA, USA

5 Harvard Medical School, Harvard University, Boston, MA, USA

6 Physical chemistry department, Complutense University of Madrid.

7 Centro de Investigaciones Biológicas Margarita Salas, Madrid, Spain

*Correspondence Sebastian Thompson, Email: sebastian.thompson@imdea.org;

|| These authors contributed equally to this work.

Abstract

Based on the PubMed database, there are around 260 manuscripts describing nanothermometers. These research articles detail the synthesis, performance, and application of intracellular nanothermometers. This intracellular prevalence is due to the significant importance, complexity, and utility of the intracellular compartments for understanding cell metabolism and disease treatment. However, in recent years, the extracellular environment of the cell has emerged as a crucial factor in medicine, particularly in hyperthermia and immunotherapy. Despite this, we have not seen evidence in the literature describing the utilization or performance of a nanothermometer designed for extracellular temperature measurements. This oversight not only neglects the potential for measuring extracellular temperature but also fails to address the extracellular environment of the cell. Here, we introduce a nanothermometer designed specifically for measuring extracellular temperature by directly converting serum proteins into nanothermometers (either unmodified or labeled with the clinically approved dye Fluorescein). Additionally, leveraging the extracellular localization of these nanothermometers, we demonstrate (1) the enhancement of their temperature sensitivity by combining them with gold nanorods, and (2) their capability to generate damage and disrupt the plasma membrane, thus opening the door to their use as photodynamic therapy agents. We firmly believe that these advancements represent not only a broadening of the applications of nanothermometry but also a pioneering step in showcasing the ability of nanothermometers to induce cell death.

In the past years, there has been a surge of interest within the scientific community in measuring temperature at the nanoscale[1,2]. This pursuit involves the use of molecules sensitive to temperature, commonly referred to as "nanothermometers"[2,3]. Nanothermometers have been crafted using various materials such as proteins, dyes, and nanoparticles, with their readout typically relying on fluorescent properties like intensity ratio, lifetime, or anisotropy[3]. However, it's noteworthy that all reported studies thus far have focused solely on intracellular temperature measurements. This emphasis stems from the pivotal role that the intracellular environment plays in cell metabolism, activity, and fate[4]. Furthermore, in disease treatment and diagnostics, there's a predominant reliance on drugs that traverse the plasma membrane to interact with the intracellular compartments[5].

However, there is a growing interest within the scientific community in the extracellular environment. Presently, techniques inducing cell death through hyperthermia and modulating immunotherapy are extending their reach beyond intracellular spaces[6–8]. Additionally, understanding the extracellular tumor environment has become crucial for comprehending tumor growth and the efficacy of anticancer therapies[9]. To the best of our knowledge, there hasn't been any reported instance of a nanothermometer specifically designed for extracellular temperature measurements.

Therefore, our approach hinges on extracting temperature information from the extracellular milieu of the cell: the serum. Serum, is derived from blood after clotting, and only differs from plasma due to the absence of clotting factors[10]. It serves as a rich source of nutrients, growth factors, hormones, and proteins critical for maintaining cell viability, proliferation, and functionality *in vitro*. The protein concentration in fetal bovine serum (FBS), mainly used for cell culture, typically ranges from 6 to 8 grams per deciliter (g/dL), with the majority being albumin, followed by globulins, immunoglobulins, carrier proteins, and other globulins[11–13].

Thus, the possibility of converting these serum proteins into nanothermometers will pave the way to translate nanoscale measurements to *in vivo* and clinical applications. Here, we present the methodology and the results of extracting temperature information from serum proteins by labeling them with clinically approved Fluorescein[14]. The methodology is based on measuring the Fluorescent Anisotropy Polarization (FPA) of the Serum itself, converting it into Serum-Anisotropy Based Nanothermometer (*Serum-ABNT*). Anisotropy-Based Nanothermometers (ABNTs) have been developed in the last ten years, and essentially any fluorescent dye and protein can be converted into ABNT[15–19]. Examples include DNA, green fluorescent protein, and any fluorescent-labeled protein. Moreover, ABNTs have been tested in cell culture and animal models.

We demonstrate that our results have the potential to extract information from the extracellular environment of cells. We foresee that by extending our method to tumors *in vivo*, can potentially have clinical relevance, thus, notably expanding the nanothermometry applications.

**Results and Discussion**

*2. Serum as nanothermometer*

As it was mentioned before, we obtain the temperature information from the FPA signal from the fluorescent signal. As it has been described vastly in bibliography, that the theoretical FPA is expressed by the Perrin equation[20]:

$$FPA = \frac{r0}{1+\tau_f/\Theta_r} \quad (1)$$

where r₀ is a constant delimiting anisotropy, τf is the fluorescence lifetime of the fluorophore, and Θr its rotational correlation time. For a spherical molecule, the rotational correlation time can be expressed by the Stokes-Einstein-Debye relation[20]

$$\Theta r = \frac{V\eta(T)}{kT} \quad (2)$$

where k is the Boltzmann constant, T and η the solvent temperature and viscosity, respectively, and V the hydrodynamic volume of the molecule, which is related to its hydrodynamic radius (Detailed theoretical studies can be found in SI). Based on previous published research[20,21], the temperature sensitivity of a fluorescent molecule is based on the match between their lifetime and hydrodynamic radius. We employ Fluorescein to label the Serum proteins as the dye is approved already in clinical applications[14]. NHS-fluorescein is a derivative of fluorescein isothiocyanate (FITC), commonly used to label proteins and other biomolecules for detection and visualization in biological samples[22]. NHS-fluorescein will have lower temperature sensitivity[18] due its mismatch between $\tau_f/\Theta_r$. However, when NHS-fluorescein is added to serum, it covalently binds to primary amines present on proteins within the serum, forming stable fluorescent conjugates[22]. Thus, this new complex (Fluorescein-Protein) will have a higher temperature sensitivity[18] based on the improved match between $\tau_f/\Theta_r$. Thus, we first study the ratio (Protein (Serum):NHS-Fluorescein) needed to avoid free fluorescein and its lower temperature sensitivity. For that, we mixed 1ml of commercial Serum with NHS-Fluorescein with different concentration of NHS-Fluorescein for 4 hours (see Methods). Figure 1a shows the anisotropy values of free NHS-Fluorescein and different Protein (Serum):NHS-Fluorescein:

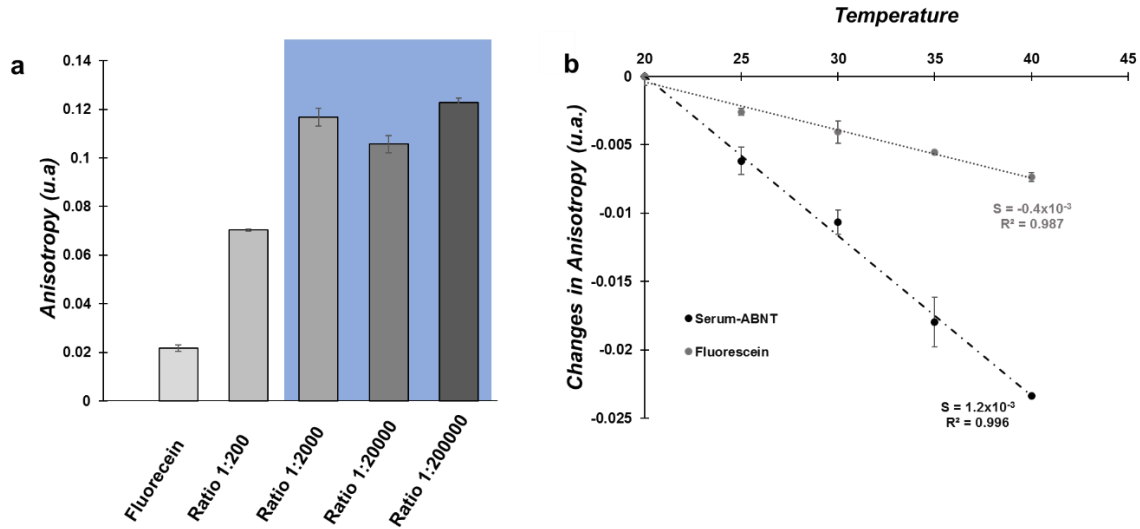

Figure 1: (a) Anisotropy in function of Protein (Serum): NHS-Fluorescein ratio. (b) Thermal sensitivity of Serum-ABNT and Fluorescein. No free Fluorescein in the light blue square ratios.

As we follow the anisotropy value of NHS-Fluorescein and Fluorescein-Proteins, NHS-Fluorescein alone present lower values of anisotropy since the dye molecule rotates faster enough to induce high light depolarization[18] (lower anisotropy). Once the fluorescein is attached to the proteins of the Serum, the complex adopts the rotation of the protein and, thus, the light polarization decreases (increase of the anisotropy values). As it can be observed in Figure 1a, ratios higher than 200 (60-80 mg protein concentration: 0.4mg/ml NHS-Fluorescein), ensure that all the fluorescein is attached to serum proteins (light blue square). For this reason, we adopt 0.04mg/ml NHS-Fluorescein:1ml Serum for the following experiments (1:2000 approximately). This excess of protein assures the absence of free NHS-Fluorescein in the final solution. The changes in temperature sensitivity of the Fluorescein-Serum and NHS-Fluorescein are illustrated in Figure 2b. As it is expected, the temperature sensitivity increases from $3.5 \times 10^{-3}$ (free fluorescein) to $1.2 \times 10^{-3}$ (Serum-ABNT). This Serum-ABNT temperature sensitivity that we obtained is in the same order of magnitude than that previously reported for BSA-Fluorescein ($9.45 \times 10^{-3}$)[18]. This similarity is expected since the most abundant protein in Serum is Albumin[11,12] (60% of the total protein). Thus, these results conclude that just by adding a clinically-approved fluorescent dye to the serum, the temperature information can be simply obtained without any other Serum modification.

*Extracellular localization of ABNT-Serum.*

ABNT-Serum is formed by Fluorescein conjugated to Serum proteins, majority to albumin. Since the Serum proteins are meant to remain extracellularly[23], we expect the same behavior for the ABNT-Serum. This is a notable difference with the clinical fluorescein (Sodium Fluorescein) that cross the plasma membrane and localize intracellularly[24]. We incubated overnight the cells with complete medium (10% Serum) supplemented with additional 10% Serum-ABNT, and then we performed confocal microscopy to assess ABNT-Serum localization, as shown in Figure 2.

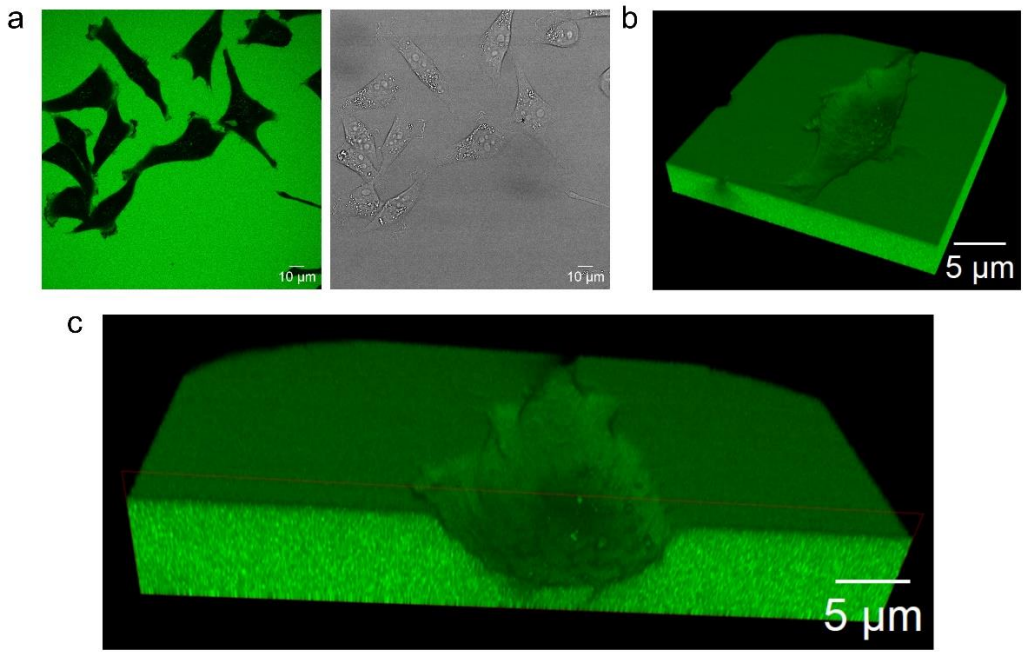

Figure 2: Extracellular localization of Serum-ABNT. (a) Representative confocal microscopy images of HeLa cells incubated with Serum-ABNT overnight. Serum-ABNT is presented in green color (left panel) and transmitted light in gray (right panel). (b) Representative 3D view of HeLa cells incubated with Serum-ABNT overnight. (c) Representative vertical section of the 3D reconstruction corresponding to the cell presented in panel b.

As expected, images of the middle plan of the cells (left green panel) show green fluorescence surrounding the cells, supporting our hypothesis that the ABNT-Serum exhibits an extracellular localization, and it is not uptake by the cells (Figure 2a). Moreover, we acquired z-stacks series covering all the cell volume using a higher magnification factor. As result, we generated 3d reconstructions that further prove the ABNT-Serum extracellular localization (Figure 2b), also applying a transversal section of the cell body to highlight the intracellular compartment (Figure 2c).

### *Increasing Serum-ABNT sensitivity by Enhanced plasmon polarization*

To increase the sensitivity of our nanothermometer, we have used gold nanorods as enhancers. As it has been reported before, gold nanoparticles have been used to increase the sensitivity of another anisotropy-based sensor[25–28]. This previous report it is based on the idea of the interaction between the anisotropy-sensitive molecule and the gold nanoparticle. Thus, we study the possibility of increasing the temperature sensitivity of the Serum-ABNT relying on the interaction of gold nanoparticles and the serum-ABNT. By simply mixing CTAB-stabilized gold Nanorods (GNRs) with the Serum-ABNT we observed their aggregation (the BSA induce the aggregation of the positively charged GNRs rather the formation of the protein Corona covering them[29]) (Figure 3a). Here we show the aggregation of the gold nanoparticles by the Serum-ABNT evidenced by the widening of the longitudinal plasmon band centered at around 800 nm[30] (Figure 3b), the changes in anisotropy and polarization values of Serum-ABNT and GNR-Serum-ABNT (Figure 3c) and the changes in temperature sensitivity of the Fluorescein, Serum-ABNT and GNR-Serum-ABNT (Figure 3d).

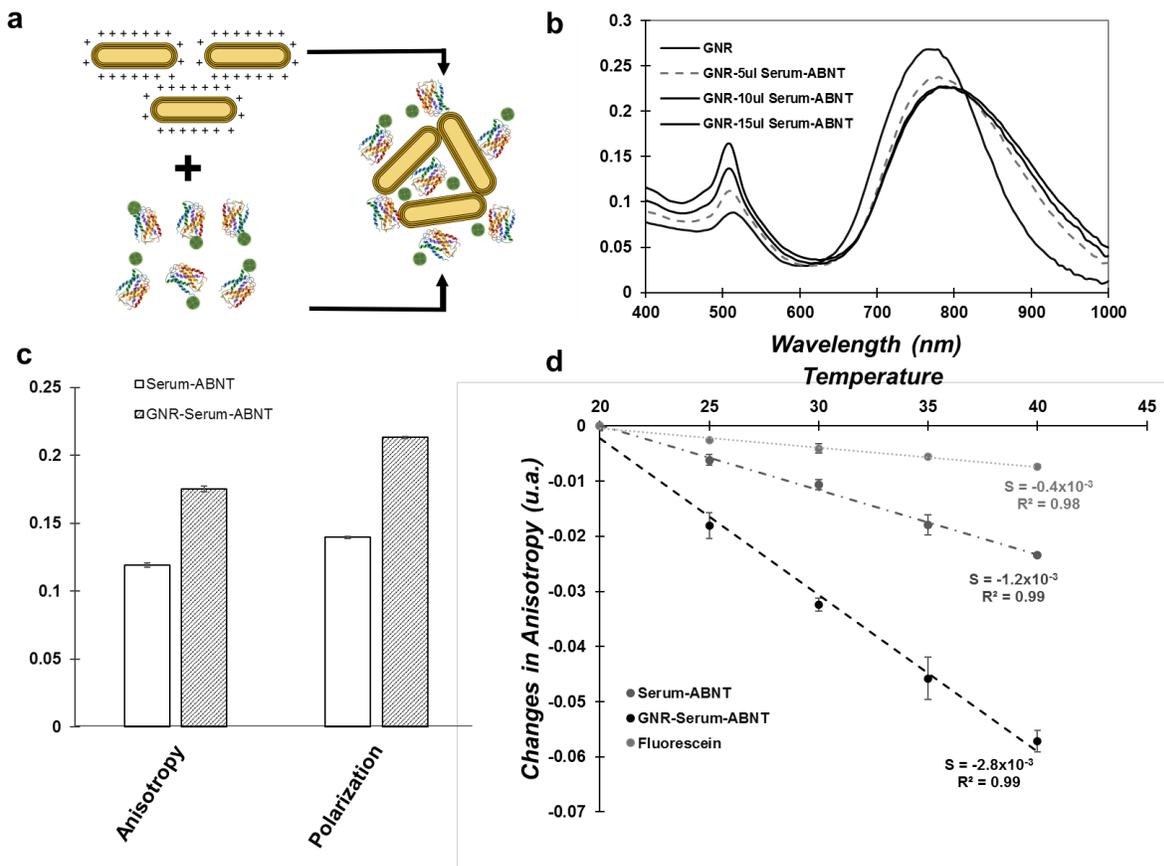

Figure 3: (a) Aggreagation of GNR by the Serum-ABNT, (b) aggregation changes measured by UV-Vis spectrophotometry, (c) changes in Anisotropy and Polarization values and (d) temperature sensitivity of Fluorescein, Serum-ABNT and GNR-Serum-ABNT.

As it was expected, the combination of GNR with the Serum-ABNT notably increase the anisotropy and polarization (defined in SI) values and thus the temperature sensitivity. The temperature sensitivity that we obtained here is $(2.8) \times 10^{-3}$ that is, interestingly, similar to the highest temperature sensitivity that can be reached by theoretical predictions[18].

*Serum-ABNT as photodynamic therapy agent*

In the last five years, there is an increasing interest in inducing cell death using photo-therapy agents located extracellularly[6,8]. When the photothermal therapy agents are located extracellularly, they compromise the stability of the plasma membrane[8,31]. Comparing with intracellular location, the extracellular location requires no longer time of incubation and the cell death mechanism is necrosis[8,31]. Thus, since Serum-ABNT is located outside of the cell, we study the possibility of damaging the plasma membrane by bleaching a restricted spot through the application of a focused point high-power laser (represented by the red dot in Figure 4).

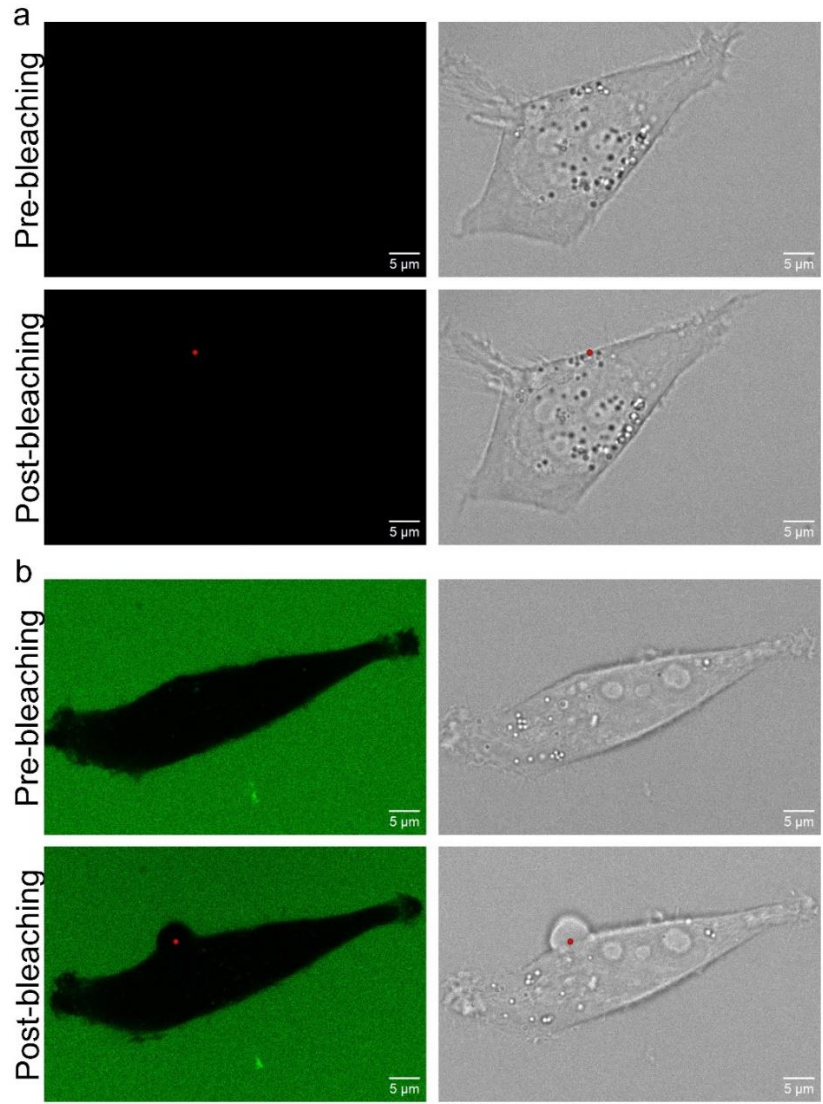

Figure 4: Plasma membrane impairment upon bleaching in HeLa treated with Serum-ABNT. Representative confocal microscopy images of (a) control or (b) serum-ABNT treated HeLa cells before bleaching (upper panel) or after bleaching (bottom panel). Serum-ABNT channel is presented in green color (left panel) and transmitted light in gray (right panel). Bleaching points are highlighted with red dots.

Images clearly demonstrate that the presence of Serum-ABNT combined with laser irradiation induce a plasma membrane impairment, as can be observed by a bubble structure formation appreciable both in the green fluorescence channel and in the transmitted light image (Figure 4b). Conversely, the control cell, cultured with medium without Serum-ABNT, does not show any membrane impairment once exposed to the same laser exposure conditions (Figure 4a), supporting the fact that Serum-ABNT can induce membrane damaging. Another control in which serum-ABNT-treated HeLa cells are irradiated with a different laser wavelength (550 nm in the red excitation spectrum) do not present any plasma membrane damaging (see supplementary Fig. SI1-middle panel). Notably, irradiating the same region with the green laser, as described previously, does induce again plasma membrane damaging, further proving that specific laser-induced Serum-ABNT excitation is necessary for cell damage (see supplementary Fig. SI1-bottom panel).

Post bleaching images showing the plasma membrane impairment presented in Figure 4 were acquired immediately after laser pulse, thus we wanted to prove that cell death was induced. To assess this, we repeat the same experiment taking images 10 and 20 minutes after bleaching, as described in Figure 5.

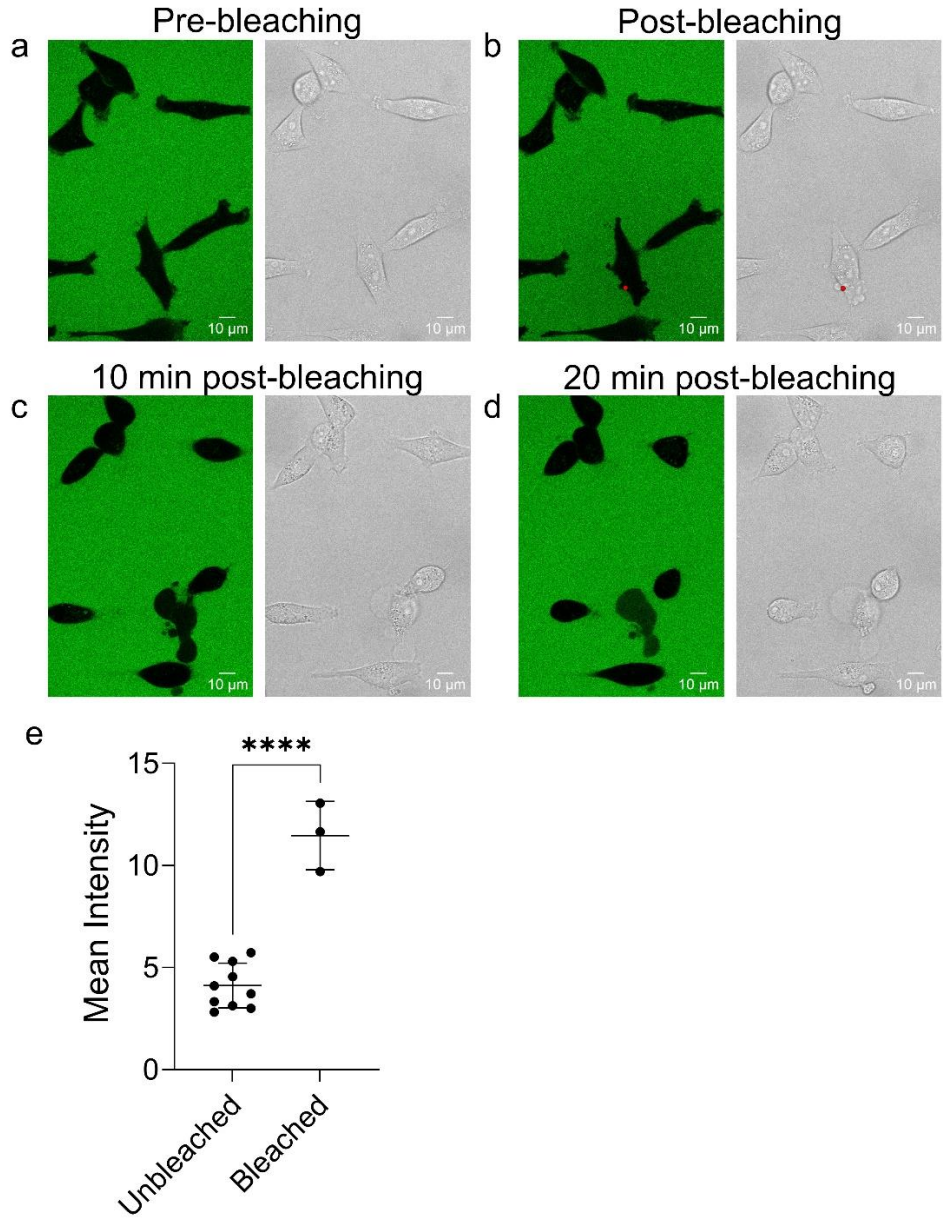

Figure 5. Cell death upon membrane impairment and fluorescein entry. Representative confocal microscopy images serum-ABNT treated HeLa cells (a) before bleaching, (b) after bleaching, (c) 10 minutes or (d) 20 minute upon bleaching. Serum-ABNT channel is presented in green color (left panel) and transmitted light in gray (right panel). Bleaching points are highlighted with red dots. (e) Mean fluorescence intensity quantification of unbleached or bleached cells 20 minutes upon bleaching performed with the Leica LAS X software designing specific region of interest delimitating single cells. Data are representative of one experiment from three different field of views. **** P value < 0.0001 assessed by unpaired t test using the GraphPad Prism 10.2.2 software.

As it can be appreciated in the Figure 6b there is a selective plasma membrane impairment by the combination of focused high-power laser and Serum-ABNT, confirming our previous results described in Figure 4. The controls (Serum-ABNT without high power bleaching) do not present any plasma membrane damaging (Figure 5a). Of note, through this technique damage is only induced in irradiated cells, and in a restricted spot of the plasma membrane. As expected, images acquired 10 and 20 minutes after the irradiation clearly show a complete damaged structure of the cell due to cell death program (Figure 6c and Figure 5d). Moreover, we could observe Serum-ABNT intracellular uptake 20 minutes after membrane damaging (Figure 5d). The difference of the intracellular mean fluorescence intensity of irradiated cells was significantly higher compared to the non-irradiated counterparts (Figure 5e).

**Discussion**

The development of Serum-ABNTs marks a significant step forward in extracellular temperature measurement. By labeling serum proteins with NHS-fluorescein, a stable and sensitive fluorescent conjugate is formed. This method effectively converts serum into a functional nanothermometer by just adding the reactive NHS-Fluorescein to the Serum[22] without extensive modifications. The resultant Serum-ABNTs exhibit similar temperature sensitivity compared with BSA-Fluorescein[18] affirming the Fluorescein attached to the BSA protein (most abundant protein in Serum[11,12]). This approach not only simplifies the creation of extracellular nanothermometers but also ensures their compatibility with clinical applications. Enhancing the thermosensitivity of Serum-ABNTs through the interaction with GNRs further amplifies their utility. The interaction between Serum-ABNTs and GNRs results in increased anisotropy and polarization values[25–28], thereby improving temperature sensitivity. The combined system reaches the highest theoretical sensitivity predicted[18], making it a powerful tool for precise temperature monitoring. This increased sensitivity can significantly impact applications in hyperthermia treatment and other therapeutic interventions where heat is generated by the GNR and the temperature control is paramount.

Confocal microscopy studies confirm the extracellular localization of Serum-ABNTs. When HeLa cells were incubated with Serum-ABNTs, the fluorescent signal was observed surrounding the cells, indicating that these nanothermometers do not penetrate the plasma membrane, something expected since Albumin and most Serum's proteins are knowing to remain outside the cells[23]. This extracellular localization is crucial for applications targeting the extracellular matrix or tumor microenvironment, where precise temperature measurements can also improve theranostic strategies. The extracellular stability of Serum-ABNTs, combined with their high sensitivity, underscores their potential for real-time monitoring in various biological and clinical settings.

The combination of Serum-ABNTs with focused laser irradiation demonstrates their capability to disrupt the plasma membrane, presenting a novel approach for photodynamic therapy[31]. When subjected to high-power laser irradiation, cells treated with Serum-ABNTs showed significant membrane damage, as evidenced by bubble formation. This effect was absent in control cells or cells irradiated with red-laser, highlighting the specific interaction between the laser and the Serum-ABNTs. This membrane-disrupting property could be leveraged for targeted cell death in cancer treatment, where precise control over the affected area is critical. The ability of Serum-ABNTs to act as both temperature sensors and therapeutic agents opens new avenues for integrated diagnostic and treatment modalities.

**Conclusions**

In conclusion, this study introduces a pioneering approach to extracellular nanothermometry by harnessing the inherent properties of serum proteins. We have demonstrated the feasibility of utilizing serum proteins as nanothermometers, both in label-free form and through conjugation with clinically approved fluorescent dyes. Label free temperature measurements can now be applied to experiments in cuvette of cell in suspension. On the other hand, we have converted the protein's Serum in nanothermometer using a clinical-approval dye. Together with this, we increase their temperature sensitivity by combining them with GNRs. We reach, in this way, the highest temperature sensitivity theoretically published[18] for ABNTs. This complex GNR-Serum-ABNT are perfect candidate to be employed in photothermal therapies, where the heat is generated by the interaction between light and GNR and the Serum-ABNT can report the changes in temperature simultaneously.

As an important application, we also report the possibility of damaging the plasma membrane utilizing the Serum-ABNT. *In vivo*, we expect the Serum-ABNT reach the tumor since the vascular permeability of BSA into solid tumor tissue[32]. Thus, the extracellular localization of Serum-ABNT will allow the measuring of the tumor temperature and the possibility of inducing tumor cell death by damaging the plasma membrane. Thus, this nanothermometer have the possibility of reporting simultaneously the temperature of tumors and damaging the plasma membrane to induce cell death.

**Methods**

Reagents

Dulbecco's modified Eagle medium (DMEM)–Glutamax (31966 057), penicillin–streptomycin (15070-063), and DMEM without phenol red (21063-029) were purchased from GIBCO. Fetal bovine serum (SV30160.3) from HyClone was heat inactivated at 56 °C for 30 min before use. The dishes for microscope imaging were 35 mm culture dishes with 20 mm treated glass bottom surface (734 2904, VWR). CTAB-stabilized gold nanorods were purchased from Nanopartz (USA). The following proteins were purchased from Sigma Aldrich: immunoglobulin G I5506, human albumin A3782, bovine serum albumin A7030 and dissolved in 1X PBS at a concentration of 1 mg/ml.

Cell Culture Preparation

Human epithelioid cervix carcinoma (HeLa) cells were maintained in DMEM–Glutamax medium (31966 057, Gibco), supplemented with 10% fetal bovine serum (SV30160.3, Hyclone) and penicillin–streptomycin (15070-063, Gibco), at 37 °C, 5% $CO_2$ atmosphere. For the experiments, 35 000 cells were plated on 35 mm culture dishes with 20 mm treated glass bottom surface (734 2904, VWR), in a final volume of 2 mL of growing medium. 24 h after, the medium was replaced by complete medium (Serum 10%) or complete medium + 10% Serum ABNT.

Serum-ABNT and GNR-Serum-ABNT synthesis

NHS-Fluorescein (Sigma) was dissolved in DMSO to final concentration of 10mg/ml. Different NHS-Fluorescein volume (40ul to 0.04ul 10mg/ml) was added to 1ml Serum (60-80 mg/ml protein concentration). The final DMSO concentration in serum was between 0.4% and 0.0004%. For the cell culture experiments, the final DMSO concentration was 0.04%. For GNR aggregation, 500ul of GNRs (6.8e11nps/ml) were diluted in 1ml PBS and different volumes of Serum-ABNT (1:2000 Ratio) was added to induce aggregation.

Confocal microscopy.

Cell were plated as described previously in the cell culture preparation paragraph of this material and methods section. Images were acquired using a Leica STELLARIS 8 STED 3X multispectral confocal system (Leica Microsystems) equipped with an incubation system (Okolab) to maintain living cells under standard culture conditions (37 °C, 5% $CO_2$), and using a water immersion HC PL APO CS2 63x NA 1.20 objective (Leica Microsystems). Samples excitation was achieved using a White laser line (WLL) settled at 488 nm with a power of 101 nW measured at the sample plane, and fluorescence emission was collected using a hybrid spectral (HyD S1) detector in the range of 500 nm-672 nm, settled in counting mode with fixed 2.5% gain factor. Transmitted light images were acquired using a Trans PMT and gain settled at 24.1%, using the same 488 nm laser. For experiments showed in Figure 3a and Figure 6, pixel size was 180 nm, and single images focusing the median cell plane have been acquired. For experiments showed in Figure 3b and Figure 3c, pixel size was 60 nm, and z stack series using a system optimize Z step size of 356 nm have been used. 3D image reconstruction was performed using the 3D package of the LAS X software, and clipping processing has been used for 3D section views. For experiments showed in Figure 4, pixel size was 75 nm, and single images focusing the median cell plane have been acquired. Bleaching condition was achieved using the WLL in the bleach point mode settled at 488 nm or 550 nm with a power of 4.5 µW or 8.30 µW respectively, measured at the sample plane, for 1 minute. In order to visualize fluorescein internalization, images were acquired at 10 and 20 minutes after bleaching.

FPA-Temperature Solution Measurements

For temperature sensitivity measurements in solution, a calibration curve relating fluorescence polarization anisotropy (FPA) to temperature was measured using a Horiba Fluorolog fluorometer running in T-format mode with vertically polarized excitation. The temperature sensitivity of each channel was corrected with horizontally polarized excitation. For each experiment, excitation and emission slits were selected to obtain around $10^5$ counts per seconds (cps). All samples were measured using a quartz cuvette. Excitation and emission monochromators were set at 497 nm and 512 nm, respectively.

**Acknowledgment**

This work was financed by the AECC (Spanish Association Against Cancer) IDEAS21989THOM. C.A.C. thanks Comunidad Autónoma Madrid for the Ayudante de investigación contract number 45576.

**References**

[1] K. M. McCabe, M. Hernandez, *Pediatr. Res. 2010 675* **2010**, *67*, 469.

[2] J. Zhou, B. del Rosal, D. Jaque, S. Uchiyama, D. Jin, *Nat. Methods* **2020**, *17*, 967.

[3] M. Quintanilla, L. M. Liz-Marzán, *Guiding Rules for Selecting a Nanothermometer*, **2018**, pp. 19:126-45.

[4] J. Zhu, C. B. Thompson, *Nat. Rev. Mol. Cell Biol.* **2019**, *20*, 436.

[5] M. F. Brana, M. Cacho, A. Gradillas, B. de Pascual-Teresa, A. Ramos, *Curr. Pharm. Des.* **2005**, *7*, 1745.

[6] G. Hannon, A. Bogdanska, Y. Volkov, A. Prina-Mello, *Nanomaterials* **2020**, *10*.

[7] R. W. Lentz, M. D. Colton, S. S. Mitra, W. A. Messersmith, *Mol. Cancer Ther.* **2021**, *20*, 961.

[8] C. Ayala-Orozco, D. Galvez-Aranda, A. Corona, J. M. Seminario, R. Range, J. N. M. L, J. M. Tour, *Nat. Chem.* **2024**, *16*, 456.

[9] A. R. Lim, W. K. Rathmell, J. C. Rathmell, *Elife* **2020**, *9*, 1.

[10] A. Vignoli, L. Tenori, C. Morsiani, P. Turano, M. Capri, C. Luchinat, *J. Proteome Res.* **2022**, *21*, 1061.

[11] M. Leeman, J. Choi, S. Hansson, M. U. Storm, L. Nilsson, *Anal. Bioanal. Chem.* **2018**, *410*, 4867.

[12] D. P. Myatt, *Biomed. Spectrosc. Imaging* **2017**, *6*, 59.

[13] M. P. M. Soutar, L. Kempthorne, E. Annuario, C. Luft, S. Wray, R. Ketteler, M. H. R. Ludtmann, H. Plun-Favreau, *Autophagy* **2019**, *15*, 2002.

[14] L. M. Wang, M. A. Banu, P. Canoll, J. N. Bruce, *Front Oncol.* **2021**, *11*, 1.

[15] J. S. Donner, S. A. Thompson, C. Alonso-Ortega, J. Morales, L. G. Rico, S. I. C. O. Santos, R. Quidant, *ACS Nano* **2013**, *7*, 8666.

[16] G. Spicer, S. Gutierrez-Erlandsson, R. Matesanz, H. Bernard, A. P. Adam, A. Efeyan, S. Thompson, *J. Biophotonics* **2020**.

[17] G. Spicer, A. Efeyan, A. P. Adam, S. A. Thompson, **2019**, 1.

[18] S. A. Thompson, I. A. Martinez, P. H.- González, A. Adam, D. Jaque, J. B. Nieder, R. De Rica, **2018**.

[19] P. Rodríguez-Sevilla, G. Spicer, A. Sagrera, A. Adam, A. Efeyan, D. Jaque, S. A. Thompson, *Adv. Opt. Mater.* **2023**, *11*.

[20] G. Baffou, C. Girard, R. Quidant, *Phys. Rev. Lett.* **2010**, *104*, 1.

[21] J. S. Donner, S. A. Thompson, M. P. Kreuzer, G. Baffou, R. Quidant, *Nano Lett.* **2012**, *12*, 2107.

[22] E. A. Berg, J. B. Fishman, *Cold Spring Harb. Protoc.* **2019**, *2019*, 229.

[23] G. L. Francis, *Cytotechnology* **2010**, *62*, 1.

[24]  M. M. Bakkar, L. Hardaker, P. March, P. B. Morgan, C. Maldonado-Codina, C. B. Dobson, *PLoS One* **2014**, *9*.

[25]  G. Wang, C. Shao, C. Yan, D. Li, Y. Liu, *J. Lumin.* **2019**, *210*, 21.

[26]  B. C. Ye, B. C. Yin, *Angew. Chemie - Int. Ed.* **2008**, *47*, 8386.

[27]  X. Wang, M. Zou, H. Huang, Y. Ren, L. Li, X. Yang, N. Li, *Biosens. Bioelectron.* **2013**, *41*, 569.

[28]  B. C. Yin, P. Zuo, H. Huo, X. Zhong, B. C. Ye, *Anal. Biochem.* **2010**, *401*, 47.

[29]  G. Wang, C. Yan, S. Gao, Y. Liu, *Mater. Sci. Eng. C* **2019**, *103*, 109856.

[30]  A.-G. Rubén, G. González-Rubio, J. G. Izquierdo, L. Bañares, I. López-Montero, A. Calzado-Martín, Montserrat Calleja, G. Tardajos, A. Guerrero-Martínez, *ACS Omega* **2016**, *1*, 388.

[31]  S. A. Thompson, A. Aggarwal, S. Singh, A. P. Adam, J. P. C. Tome, C. M. Drain, *Bioorganic Med. Chem.* **2018**, *26*, 5224.

[32]  H. Cho, S. I. Jeon, C. H. Ahn, M. K. Shim, K. Kim, *Pharmaceutics* **2022**, *14*.